\documentclass[pra,twocolumn,floatfix,superscriptaddress,showpacs,nofootinbib]{revtex4}

\usepackage{amsfonts}
\usepackage{amssymb}
\usepackage{amsmath}
\usepackage{graphics}
\usepackage{epsfig}

\usepackage{color}

\newcommand{\ket}[1]{{|{#1}\rangle}}
\newcommand{\tmin}{\theta_{\mbox{{\tiny min}}}}

\begin{document}

%-----------------------------------------------------------------------------

\title{Entanglement is not a critical resource for quantum metrology}
\author{Todd Tilma}\email{ttilma@nii.ac.jp}
\affiliation{National Institute of Informatics, 2-1-2 Hitotsubashi, Chiyoda-ku, Tokyo 101-8430, Japan}
\author{Shinichiro Hamaji}
\affiliation{Department of Physics, The University of Tokyo, Tokyo 113-0033, Japan}
\affiliation{National Institute of Informatics, 2-1-2 Hitotsubashi, Chiyoda-ku, Tokyo 101-8430, Japan}
\author{W. J. Munro}
\affiliation{Hewlett Packard Laboratories, Bristol BS34 8HZ, United Kingdom}
\affiliation{National Institute of Informatics, 2-1-2 Hitotsubashi, Chiyoda-ku, Tokyo 101-8430, Japan}
\author{Kae Nemoto}
\affiliation{National Institute of Informatics, 2-1-2 Hitotsubashi, Chiyoda-ku, Tokyo 101-8430, Japan}

\date{\today}

\begin{abstract}
We have investigated high-precision measurements, beyond the standard quantum limit, utilizing non-classical states.  
Although entanglement has been considered a resource for achieving the Heisenberg limit in measurements, we show that any advantage expected from using entanglement is dependent on the measurement in question.  
We investigate several measurement scenarios and illustrate the role of entanglement as a resource for quantum high-precision measurement.  
In particular, we demonstrate that there is a regime wherein entanglement not only does not help, but prevents the achievement of the fundamental limit.
\end{abstract}

% 03.65.Ta 	Foundations of quantum mechanics; measurement theory
% 03.65.-w 	Quantum mechanics
% 03.67.-a 	Quantum information
% 42.50.Dv 	Quantum state engineering and measurements
% 42.50.Xa 	Optical tests of quantum theory

\maketitle
%
%-----------------------------------------------------------------------------
\section{Introduction\protect}\label{Introduction}

Quantum metrology is a field that investigates characteristic fundamental properties of measurements under the laws of quantum mechanics~\cite{DowlingReview}. 
Quantum high-precision measurements, in particular, focuses on realizing more precise measurements, measurements beyond their classical counterparts. 
They have the ultimate goal of achieving the informatic bounds allowed by the laws of quantum mechanics.
The typical and classic example is measurements utilizing squeezed states.
It is well known that squeezed states exhibit sensitivities beyond the standard quantum limit~\cite{Caves1981} both theoretically and experimentally.  
Squeezed states can approach the Heisenberg limit with infinite squeezing, which indicates the infinite amount of energy necessary to achieve the fundamental quantum limit.  

In recent years, quantum information science has rapidly developed, and quantum measurement schemes, as information gathering processes, have been widely investigated from a quantum information point of view.  
The development of experimental quantum information processing (QIP)~\cite{Cory1994,CiracZoller1995,Gershenfeld1997,NEC1999,NJP2006,Kok2007,alleaume-2007} has ignited a strong motivation to realize quantum measurements beyond the squeezed-state regime and to test the ultimate limits for measurements.  
The fundamental technologies necessary for QIP have led us to this new stage of quantum metrology research where entanglement plays a natural role.  

Now that non-classical states, typically entangled states, are available in laboratories, a number of protocols to implement quantum metrology have been considered and several of them have actually been implemented~\cite{Cory1994,CiracZoller1995,Gershenfeld1997,NEC1999,NJP2006,Kok2007,alleaume-2007}.  
Given that entanglement has a central role in most QIP protocols, naturally, QIP developments invoked strong interest in whether the theory for non-classicality including entanglement could contribute to investigations of quantum high-sensitivities.  
Entangled photon number states have been shown to help phase estimation and lithographic resolution beat the standard quantum limit~\cite{Boto,Kok}. 
Also, more general entangled light states have been shown to greatly improve photodetector calibration~\cite{Migdall}, quantum imaging~\cite{Dowling2006}, and lidar~\cite{Lloyd}. 
In all of these successful protocols, entanglement has indeed played the central role in achieving super-sensitivity below the shot-noise limit~\cite{DowlingReview}.  
These results cultivated a belief that entanglement is essential to the mechanism to allow us to access the Heisenberg limit with a finite amount of energy.

The main aim of this paper is to challenge this belief.  
Entanglement is a key resource for a large number of QIP applications and there are a number of indications that entanglement is the necessary element for achieving high-precision measurements~\cite{boixo:040403,boixo:090401}. 
However, there are somewhat different views about the role of entanglement in quantum mechanical and quantum computational processes~\cite{PRA1,PRA2}. 
For instance, the Gottesman-Knill theorem was the first to show that entanglement alone is not sufficient for universal quantum computation~\cite{Gottesman99,Bartlett02}.  
Although entanglement is necessary for scalable universal quantum computation~\cite{Jozsa2003}, scalability might not necessarily be applicable to quantum metrology.   

In this paper, we demonstrate how the fundamental limits on sensitivity can be achieved under the principles of quantum mechanics.  
The scaling of the sensitivities approaching the Heisenberg limit will be investigated using several characteristic quantum states.  
We employ several states and Hamiltonians to illustrate the characteristics of the fundamental measurement limits, however the discussion applies to a broad class of states and the main results hold in general.
%
%-----------------------------------------------------------------------------
\section{The Model\protect}\label{Model}

To begin, we consider a physical parameter $\theta$ to be estimated, which is associated with the evolution represented by the unitary transformation
\begin{eqnarray}
U(\theta) = e^{i\theta\hat{H}/\hbar},
\label{WeakUnitary}
\end{eqnarray}
where $\hat{H}$ is the effective Hamiltonian of the system.  
We assume $\hat{H}= \sum_{i=1}^{N}{\hat{H}}_i = \hbar \sum_{i=1}^{N}  \hat{n}_i^k$, where $N$ is the number of subsystems involved, $\hat{n}_i= \hat{a}_i^\dagger \hat{a}_i$ is the number operator for subsystem $i$, and $k$ is the order parameter of the non-linearity.
$k=1$ corresponds to a linear phase shift on the state of the system and $k=2$ gives the well known nonlinear Kerr phase shift~\cite{Alfano:86}.  
We can also consider non-integers for $k$, such as $k= 1 / 2$, which represents a square-root-type phase shift useful for the preparation of Schr\"odinger-cat-like states~\cite{Alexei2}.  

Equation~\eqref{WeakUnitary} can be written in the form 
\begin{eqnarray}
U(\theta) = U_1 (\theta) \otimes U_2 (\theta) \otimes \ldots \otimes U_N (\theta)
\label{UnitaryExpansion}
\end{eqnarray}
as each ${\hat{H}}_i$ acts only on its own subspace.  
We assume this non-entangling operation, and the no-cost resource on the evolution, so as to first demonstrate the role of entanglement in high-precision measurements.  
However, more generally, the Hamiltonian could be an entangling interaction and the unitary transformation might be realized with a resource cost.  
We will later extend our discussion beyond these assumptions.

To evaluate the parameter $\theta$, we first prepare a probe in a quantum state $\ket{\psi}$.  
Due to the assumption of no-cost resource on the unitary transformation, we do not need to consider strategies for efficient use of the physical resource to generate the unitary transformation, the state simply evolves via (\ref{WeakUnitary}) to a state $\ket{\psi'} = U(\theta) \ket{\psi}.$
Using $\ket{\psi'}$ we can estimate the smallest $\theta$ that can be detected/resolved with the following Cramer-Rao uncertainty relation~\cite{SamCaves94,general_estimation_theory}
\begin{eqnarray}
\bigg\langle \bigg(  \frac{\theta_{\rm est}}{\partial\langle \theta_{\rm est} \rangle_\theta / \partial \theta }- \theta \bigg)^{\!2} \bigg\rangle\geq \frac{\partial \theta^{2}}{4M(1-d)},
\label{othersamcaves}
\end{eqnarray}
where $d = |\langle{\psi}|{\psi'}\rangle|^2$ is the distinguishability and $M$ the number of classical measurements employed. 
If $d \sim 0$ (i.e.~the states $\ket{\psi}$ and $\ket{\psi'}$ are effectively orthogonal) then the inequality is most easily satisfied, and so $d \sim 0$ will be our metric here.

For illustrative purposes we employ several initial states $\ket{\psi}$ to highlight the effect of entanglement on the distinguishability $d$.  
The initial states undergo the unitary transformation given by $U(\theta)$, where $\theta$ is the parameter that indicates the sensitivity.  
The first choice of initial states are several, distinct, non-classical states
\begin{eqnarray}
\ket{\psi}_{C} &=&\frac{1}{\sqrt{2}}\biggl (\ket{0} + \ket{\sqrt{N} \alpha}\biggr),
\label{coherentcat} \\
\ket{\psi}_{E} &=&\frac{1}{\sqrt{2}}\biggl (\ket{0}_1\cdots \ket{0}_N + \ket{\alpha}_1 \cdots \ket{\alpha}_N \biggr),
\label{coherent_one} \\
\ket{\psi}_{S} &=& \frac{1}{\sqrt{2^N}}\biggl (\ket{0} + \ket{\alpha}\biggr)^{\otimes_N},
\label{coherent_three}
\end{eqnarray} 
where $\alpha^2 \gg 1$ for the state normalization.
These three states typically exhibit different superposition and entanglement properties, nonetheless each of these states has the same mean photon number, $\bar{n} = N|\alpha|^2 / 2$, which gives the same energy constraint on these states~\cite{footnote1,footnote2}. 

Now, lets see the effect of~\eqref{WeakUnitary} on the initial states (\ref{coherentcat}-\ref{coherent_three}). 
This is best seen through the calculation of the distinguishability for each state
\begin{eqnarray}
d_{C} &\sim&\frac{1}{2}\biggl (1 + \cos \left[\theta (N |\alpha|^2)^k \right] \biggr), 
%\ket{\psi'}_{C} &\sim&\frac{1}{\sqrt{2}}(\ket{0} + e^{i\theta (N |\alpha|^2)^k}\ket{\sqrt{N} \alpha}),
\label{coherentcatout} \\
d_{E} &\sim&\frac{1}{2}\biggl (1 + \cos \left[\theta N |\alpha|^{2k} \right] \biggr), 
%\ket{\psi'}_{E} &\sim&\frac{1}{\sqrt{2}}(\ket{0}_1\cdots \ket{0}_N + e^{i\theta N |\alpha|^{2k}} \ket{\alpha}_1 \cdots \ket{\alpha}_N),
\label{coherent_oneout} \\
d_{S} &\sim&\frac{1}{2^N}\biggl (1 + \cos \left[\theta |\alpha|^{2k} \right] \biggr)^N,
%\ket{\psi'}_{S} &\sim& \frac{1}{\sqrt{2^N}}(\ket{0} + e^{i\theta |\alpha|^{2k}}\ket{\alpha})^{\otimes N},
\label{coherent_threeout}
\end{eqnarray} 
where we have assumed that $\theta^2 |\alpha|^{2k}\ll 1$~\cite{Ralph2001,master}.
%where, for the sake of simplicity, we have employed the approximation $e^{i\theta \hat{n}^k} \ket{\alpha}\sim e^{i\theta |\alpha|^{2k}} \ket{\alpha}$ for $\theta^2 |\alpha|^{2k}\ll 1$ with $\alpha \gg 1$~\cite{Ralph2001,master,footnote4}. 
By setting $d=\delta \ll 1$ we may determine the sensitivity for $\theta$ in each case.
%As the distinguishability for each case can be calculated as the overlap between the initial state and the final state, by setting $d=\delta \ll 1$ we may determine the sensitivity for $\theta$ \cite{footnote3}. 
For instance, the distinguishability between $\ket{\psi'}_{C}$ and $\ket{\psi}_{C}$ gives 
\begin{eqnarray}
\theta_{C} \sim \frac{\cos^{-1} \left[ 2 \delta -1\right]}{\left (N|\alpha|^{2}\right)^k} \sim \frac{\left[ (2 l+1)\pi - 2 \sqrt{\delta} \right]}{ \left (N |\alpha|^{2}\right)^k},
\label{DistinguishabilityCalculation}
\end{eqnarray} 
where $l$ is a non-negative integer. 
Setting $l=0$ gives the first minimum, which corresponds to the highest precision the scheme could provide.  
Applying this line of reason to the other cases, and recalling $\bar{n} = N |\alpha|^2/2$, we have
%The minimum theta ($\tmin$) for all three cases is therefore
\begin{eqnarray}
{\tmin}_{C} &\sim& \frac{\pi - 2 \sqrt{\delta} } {(2 \bar{n})^k},
\label{coherentcattheta} \\
{\tmin}_{E} &\sim& \frac{(\pi- 2 \sqrt{\delta}) N^{k-1}} {(2 \bar{n})^k},
\label{coherent_onetheta} \\
{\tmin}_{S} &\sim& \frac{(\pi- 2 \sqrt{\delta}) N^{k-\frac{1}{2}}} {(2 \bar{n})^k}.
\label{coherent_threetheta}
\end{eqnarray}
%
%-----------------------------------------------------------------------------
\section{Resolution\protect}\label{Observations}

We can now investigate the scaling for these three states with respect to the non-linear phase shifts.
Equations (\ref{coherentcattheta}-\ref{coherent_threetheta}) are scaling as $1 / \bar{n}^{k}$, as expected, but the precision limit for each scheme is critically dependent on the factors $N$ and $k$ in the coefficient.  
The precision limits comparing (\ref{coherent_onetheta}) and (\ref{coherent_threetheta}) definitely indicate that entanglement always helps. 
However, it depends on $k$ as to whether (\ref{coherentcattheta}) is better or worse than (\ref{coherent_onetheta}), that is, whether entanglement helps or not.
For the case $0<k<1$, a large $N$ is optimal. 
We obtain an improvement by dividing our given mean energy resource over distributed modes using entanglement.  
The precision given by ${\tmin}_{E}$ is the optimal choice and achieves the Cramer-Rao bound. 
This agrees with the previous results~\cite{Munro,Giovannetti2006}. 

In contrast to the scaling for $0<k<1$, the regime for $k>1$ typically shows that large $N$ actually reduces the precision.  
In this regime, entanglement does not help the precision and is a major disadvantage.  
The disadvantage from the energy distribution over multiple modes exceeds the precision gain provided by entanglement and so it is better to consider a single mode superposition of coherent states (\ref{coherentcat}). 
Such a strategy allows one to also reach the Cramer-Rao bound.  
The $k=1$ case, in between these two regimes, is where the gain and loss are balanced and there is no difference in ${\tmin}_{C}$ and ${\tmin}_{E}$.  
We summarize the scaling properties in Table~\ref{table_coherent} for the $\delta=0$ situation. 
\begin{center}
\renewcommand{\arraystretch}{2}
\begin{table}[htb]
\begin{tabular}{|c|ccc|} \hline
$k$ & ${\tmin}_{C}$ & ${\tmin}_{E} $ & ${\tmin}_{S}$ \\ \hline 
%$\hat{H}=\hat{n}^{\frac{1}{2}}$ & $\pi ({2\bar{n}})^{-\frac{1}{2}}$ & $\pi ({2 N \bar{n}})^{-\frac{1}{2}}$ & $\pi ({2\bar{n}})^{-\frac{1}{2}}$ \\
$\hat{H}=\hat{n}^{1/2}$ & $\frac{\pi}{\sqrt{2\bar{n}}}$ & $\frac{\pi}{\sqrt{2 N \bar{n}}}$ & $\frac{\pi}{\sqrt{2\bar{n}}}$ \\
%$\hat{H}=\hat{n}$ & $\pi ({2\bar{n}})^{-1}$ & $\pi ({2\bar{n}})^{-1}$ & $\pi \sqrt{N} ({2\bar{n}})^{-1}$ \\
$\hat{H}=\hat{n}$ & $\frac{\pi}{2\bar{n}}$ & $\frac{\pi}{2\bar{n}}$ & $\frac{\pi \sqrt{N}}{2\bar{n}}$ \\
%$\hat{H}=\hat{n}^{2}$ & $\pi ({4\bar{n}^2})^{-1}$ & $\pi N ({4\bar{n}^2})^{-1}$ & $\pi N^{\frac{3}{2}} ({4\bar{n}^2})^{-1}$ \\ \hline
$\hat{H}=\hat{n}^{2}$ & $\frac{\pi}{4\bar{n}^2}$ & $\frac{\pi N}{4\bar{n}^2}$ & $\frac{\pi N^{\frac{3}{2}}}{4\bar{n}^2}$ \\ \hline
%General ($\hat{H}=\hat{n}^k$) & $\pi {(2\bar{n})^{-k}}$ & $\pi N^{k-1} {(2\bar{n})^{-k}}$ & $\pi N^{k-\frac{1}{2}} {(2\bar{n})^{-k}}$  \\ 
$\hat{H}=\hat{n}^k$ & $\frac{\pi}{(2\bar{n})^k}$ & $\frac{\pi N^{k-1}}{(2\bar{n})^k}$ & $\frac{\pi N^{k-\frac{1}{2}}}{(2\bar{n})^k}$ \\ 
\hline
\end{tabular}
\caption{The minimal detectable $\tmin$ for the unitary transformation given by (\ref{WeakUnitary}) with the initial states (\ref{coherentcat}-\ref{coherent_three}).  
The general $k$ case is given, as well as the special cases $k=1/2,\;1,\;\text{and }2$. 
These show that the entangled coherent state superposition is better in the region $0<k<1$ while the single mode coherent state superposition is better for $k>1$, indicating entanglement makes the precision worse. 
For $k=1$ (the linear phase shift case) there is no real advantage from entanglement.}
\label{table_coherent}
\end{table}
\end{center}

As shown in Table~\ref{table_coherent}, the scaling properties are defined by the energy distribution and the entanglement in the states, so our results can be easily extended to different types/superpositions of coherent states.  
We can, for instance, consider the more general state 
\begin{eqnarray}
\ket{\phi}= \sum_{i_1,i_2\ldots i_N} c_{i_1,i_2\ldots i_N} \ket{\alpha_{i_1}} \ket{\alpha_{i_2}}  \dots \ket{\alpha_{i_N}}.
\label{GeneralPhi}
\end{eqnarray}   
This state does not require the even distribution of energy over the modes, and there are a large class of states that can be categorized by the same scaling.  
In fact it is straightforward to show, once the total mean energy is fixed, that the optimal $\ket{\phi}$ corresponds to (\ref{coherent_one}) for $0<k<1$ and (\ref{coherentcat}) for $k>1$. 
We also do not need to restrict our attention to superpositions of coherent states. 
We could employ superpositions of number states instead, typically giving the equivalent states $1 / \sqrt{2}(\ket{0} + \ket{N})$ and $1 / \sqrt{2}(\ket{0}_1\cdots \ket{0}_N + \ket{1}_1\cdots \ket{1}_N)$, where the same scaling properties can be obtained.  
We note that the same generalization holds with the number state as well as the coherent states (see the Appendix).

The $k>1$ situation is interesting because it places entangled resources at a disadvantage. 
We can consider possibilities where entanglement regains the capability to achieve the sensitivity limit ${\tmin}_{C}$ by using an entangling unitary transformation.  
Changing the Hamiltonian in~\eqref{WeakUnitary} from $\hat{H}= \hbar \sum_{i=1}^{N}  \hat{n}_i^k$ to $\hat{H}= \hbar \left(\sum_{i=1}^{N}  \hat{n}_i\right)^k$, 
an $N$-body interaction~\cite{boixo:040403,boixo:090401}, the entangled state case~\eqref{coherent_one} gives the sensitivity ${\tmin}=\pi / (2 \bar{n})^{k}$, the same as the one given in~\eqref{coherentcat}. 
The entangled state merely regains the same sensitivity as the non-entangling state.

Next we need to consider the scenario where there is no a priori knowledge about the magnitude of the parameter $\theta$ to estimate.  
We can evaluate the scalability with respect to the cost associated to achieve the sensitivity and compare it to the scaling on the sensitivity limit given by (\ref{coherentcattheta}-\ref{coherent_threetheta}). 
In that calculation, $\tmin$ requires $\theta$ to be restricted to the range $0 < \theta \leq \pi / \bar{n}$. 
This requires a priori knowledge that we may not have. However, a simple solution exists. 
We could start with $0 < \theta \leq 2\pi$ and use some of our resources to iteratively refine our $\theta$ region. 
This requires, at most, doubling our total mean photon number and $\text{log}_2 (\bar{n})+1$  steps. 
For instance, with our superposition of coherent states, the total mean photon number required becomes $\bar{n}_{\rm tot} = 2 \bar{n}-1$ where $\bar{n}$ is the mean photon number required for the measurement in the range $0 < \theta \leq \pi / \bar{n}$. 
This means we can easily write $\tmin$ in equations (\ref{coherentcattheta}-\ref{coherent_threetheta}) in terms of $\bar{n}_{\rm tot}$ and thus observe the scaling in terms of the total mean photon number.

So far we have focused our attention on the sensitivity when the unitary transformation acts equally on the modes.  
In some cases, such as gravitational wave detection, this is the case. However, there are applications where the total resource to generate the unitary transformation $U(\theta)$ could be limited.  
In these applications, we could potentially be using $N$ times the unitary resources of the single mode case.  
Here we extend our analysis to estimate the precision limit in such a scenario.  
Consider the situation where the total amount of unitary resource is fixed to $U(\theta)$, but is allowed to be split into smaller pieces, with each piece used only once.  
The minimum $\tmin$ for the three states of interest becomes
\begin{eqnarray}
{\tmin}_{C} &=& \frac{\pi} {(2 \bar{n})^k},\\
{\tmin}_{E} &=& \frac{\pi N^{k}} {(2 \bar{n})^k}, \\
{\tmin}_{S} &=& \frac{\pi N^{k+\frac{1}{2}}} {(2 \bar{n})^k}.
\label{restricted-case}
\end{eqnarray}
This result clearly shows that the single mode coherent state superposition given by~\eqref{coherentcat} always achieves the best sensitivity limit, independent of the value of $k$. 
Naturally, entanglement is always disadvantage. 
With constrained resources, ${\tmin}_{E}$ and ${\tmin}_{S}$ are worse than before by a factor of $N$.   

As we have mentioned, these arguments can be readily generalized.  
In particular, for application purposes, various superpositions of number states may be of interest (see the Appendix).  
For the use of number states of the type $1/ \sqrt{2}(\ket{0} + \ket{N})$ and $1 / \sqrt{2}(\ket{0}_1\cdots \ket{0}_N + \ket{1}_1\cdots \ket{1}_N)$ in the case where the unitary operation acts evenly for all modes, we have ${\tmin}=\pi / (2 \bar{n})^{k}$ for the qudit superposition and ${\tmin}=\pi / (2 \bar{n})$ for the entangled state (here $\bar{n} = N / 2$). 
In the constrained resource case, we have ${\tmin}=\pi / (2 \bar{n})^{k-1}$ and ${\tmin}=\pi $ respectively, showing that the previous argument exactly follows in the same way.  
In fact, these states has been generated in optics from microwave to optical frequencies.  
In the case where the direct creation of $\ket{0} + \ket{N}$ is difficult, we alternatively can use a ``high N00N'' state  of the form $\ket{0}_1\ket{N}_2 + \ket{N}_1\ket{0}_2$~\cite{Lee2002}.  
Although the ``high N00N'' state is entangled, we are not making use of this property during the metrology part of the operation. 
Only one of the two modes interacts with the unitary operation. 
The other mode is passive and takes not active part. 
This second mode could be measured in a basis that gives no information about the photon number components of that mode, in effect transforming the ``high N00N'' state to $\ket{0}_1 + \ket{N}_1$. 
Alternatively, it could be left there.
%
%-----------------------------------------------------------------------------
\section{Conclusion\protect}\label{Conclusion}

In conclusion, we have analyzed the fundamental limitations on high-precision measurements utilizing quantum nature. 
It is clear that non-classical states must be used to achieve the Heisenberg limit, even with finite energy.
However, entanglement is not a necessary resource for quantum high-precision measurements.  
Our analysis shows that there are scenarios where entanglement can even be a disadvantage, preventing us from achieving the fundamental sensitivity limit (that is the Heisenberg limit).  
This result is rather counter-intuitive. 
Nevertheless, when we carefully consider the distribution of resources for sensitivity, we observe that entanglement can be used to recover the disadvantage caused from the energy distribution over multiple modes to some degree. 
Yet the loss in sensitivity cannot always be recovered. 
This is not to say entanglement is not useful, it may just not be necessary during the quantum metrology process, but could be used in generating the initial quantum state resource. 
Finally, our considerations so far have been highly ideal in the sense that we have not considered loss and decoherence effects. 
Such effects might prevent our schemes from achieving the ultimate sensitivity shown in this paper and additional mechanisms may be required to compensate for it.
%
%-----------------------------------------------------------------------------
\section{Acknowledgments\protect}\label{Acknowledgments}

We thank Gerard Milburn, Samuel Braunstein, and E. C. G. Sudarshan for valuable discussions. 
This work was supported in part by MEXT, HP, and the EU project QAP.
%
%-----------------------------------------------------------------------------
\section{Appendix - Number States\protect}\label{Appendix}

Let us consider our superpositions of number states in slightly more detail. 
For
\begin{eqnarray}
\ket{\psi}_C &=& \frac{1}{\sqrt{2}}\biggl (\ket{0} + \ket{N} \biggr),
\label{fock_one}\\
\ket{\psi}_E &=& \frac{1}{\sqrt{2}}\biggl (\ket{0}_1 \cdots \ket{0}_N + \ket{1}_1 \cdots \ket{1}_N \biggr),
\label{fock_two}\\
\ket{\psi}_S &=& \frac{1}{\sqrt{2}^{N}} \biggl (\ket{0} +\ket{1} \biggr)^{\otimes N},
\label{fock_three}
\end{eqnarray}
%with mean photon number given by $\bar{n} = N / 2$.
%As before, this mean photon number is proportional to the energy of the state and hence its value gives an indication of the amount of energy resources available to us.
%
we can apply our transformation~\eqref{WeakUnitary} and determine the distinguishability $d$. 
This gives
\begin{eqnarray}
d_{C} &=& \frac{1}{2} \biggl (1 + \cos \left[\theta N^k \right] \biggr),
\label{fock_one_distiguishability}\\
d_{E} &=&\frac{1}{2}\biggl (1 + \cos \left[\theta N \right] \biggr),
\label{fock_two_distiguishability}\\
d_{S} &=&\frac{1}{2^N} \biggl (1 + \cos \left[\theta \right] \biggr)^N,
\label{fock_three_distinguishability}
\end{eqnarray}
%Note that here, unlike in the coherent state cases, our distinguishability is exact. 
%
%By setting $d=\delta \ll 1$ we may determine the sensitivity for $\theta$
and it is straightforward to show
\begin{eqnarray}
\theta{_C} &\sim& \frac{\pi }{(2 \bar{n})^k },
\label{fock_one_theta}\\
\theta_{E} &\sim& \frac{\pi }{2 \bar{n} },
\label{fock_two_theta}\\
\theta{_S} &\sim& \frac{\pi }{\sqrt{2 \bar{n}}},
\label{fock_three_theta}
\end{eqnarray}
%Here we have substituted the mean photon number $\bar{n}$ for $N / 2$.
%For~\eqref{fock_three_theta} it is obvious that we can perform this measurement $N$ times~\cite{Kok,Giovannetti2006}.
where for~\eqref{fock_three_theta} there was an additional $1/\sqrt{N} $ improvement from classical statistics.
%\begin{eqnarray}
%\theta{_S} \sim \frac{\left[ (2 l+1)\pi - 2^N \delta^{\frac{1}{N}} \right]}{\sqrt{2 \bar{n}}}.
%\label{fock_three_update_min}
%\end{eqnarray}
These results are summarized in Table~\ref{table_fock} for various $k$.
\begin{center}
\renewcommand{\arraystretch}{2}
\begin{table}[htb]
\begin{tabular}{|c|ccc|} \hline
$k$ & ${\tmin}_{C}$ & ${\tmin}_{E} $ & ${\tmin}_{S}$  \\ \hline 
$\hat{H}=\hat{n}^{\frac{1}{2}}$ & $\frac{\pi}{\sqrt{2\bar{n}}}$ & $\frac{\pi}{2\bar{n}}$ & $\frac{\pi}{\sqrt{2\bar{n}}}$ \\
$\hat{H}=\hat{n}$ & $\frac{\pi}{2\bar{n}}$ & $\frac{\pi}{2\bar{n}}$ & $\frac{\pi}{\sqrt{2\bar{n}}}$ \\
$\hat{H}=\hat{n}^{2}$ & $\frac{\pi}{(2\bar{n})^{2}}$ & $\frac{\pi}{2\bar{n}}$ & $\frac{\pi}{\sqrt{2\bar{n}}}$ \\ \hline
$\hat{H}=\hat{n}^k$ & $\frac{\pi}{(2\bar{n})^{k}}$ & $\frac{\pi}{2\bar{n}}$ & $\frac{\pi}{\sqrt{2\bar{n}}}$ \\
\hline
\end{tabular}
\caption{The minimal detectable $\tmin$ for the unitary transformation given by (\ref{WeakUnitary}) with the initial states (\ref{fock_one}-\ref{fock_three}).  
The general $k$ case is given, as well as the special cases $k=1/2,\;1,\;\text{and }2$. 
These show that the $N$-mode entangled number state is better in the region $0<k<1$ while the single mode number state is better for $k>1$, indicating entanglement makes the precision worse. 
For $k=1$ (the linear phase shift case) there is no real advantage from entanglement.}
\label{table_fock}
\end{table}
\end{center}

From Table~\ref{table_fock} we can already see that, for the phase shift case $(k = 1)$, both $N$-mode entangled number states, given by~\eqref{fock_two}, and single mode number states, given by~\eqref{fock_one}, are good to use. 
However, when we look at $k$ values greater than 1, states of the form~\eqref{fock_one} offer better resolution, at higher mean photon numbers, than states of the form~\eqref{fock_two}. 
More interestingly, when $k < 1$, the situation reverses itself. 
States of the form~\eqref{fock_two} becomes the better choice for high-precision measurements.
%What help does our $N$-mode superposition state, given by~\eqref{fock_three}, offer?
%Except in the special case $(k = 0)$, it's never a good state to use. 

Moving to the situation where we constrain both the mean photon number of the system, as well as the total resources $U(\theta)$, we find
\begin{eqnarray}
\tmin{_C}^\prime &=& \frac{\pi}{(2 \bar{n})^{k-1}}, \\ 
\tmin{_E}^\prime &=& \pi, \\ 
\tmin{_S}^\prime &=& \pi\sqrt{(2 \bar{n})}.
\label{eprime_items}
\end{eqnarray}
This clearly shows that~\eqref{fock_one} offers the best resolution for $k > 0$.
Furthermore, both the unconstrained and constrained cases show the same scaling as we saw previously for coherent states.
%Because of this transformation, for $k > 0$,~\eqref{fock_one} is now going to offer better resolution, at higher $N$ values/mean photon numbers, than~\eqref{fock_two}.
%
%-----------------------------------------------------------------------------
%
% ***** Bibliography *****
%


\begin{thebibliography}{10}

\bibitem{DowlingReview}
J.~P. Dowling, \newblock Cont. Phys. {\bf 49}, 125 (2008).

\bibitem{Caves1981}
C.~M. Caves,
\newblock Phys. Rev. D {\bf 23}, 1693 (1981).

\bibitem{Cory1994}
D.~G. Cory, A.~F. Fahmy, and T.~F. Havel,
\newblock Proc. Nat. Acad. Sci. {\bf 94}, 1634 (1994).

\bibitem{CiracZoller1995}
J.~I. Cirac and P.~Zoller,
\newblock Phys. Rev. Lett. {\bf 74}, 4091 (1995).

\bibitem{Gershenfeld1997}
N.~A. Gershenfeld and I.~L. Chuang,
\newblock Science {\bf 275}, 350 (1997).

\bibitem{NEC1999}
Y.~Nakamura, Y.~A. Pashkin, and J.~S. Tsai,
\newblock Nature {\bf 398}, 786 (1999).

\bibitem{NJP2006}
J.~L. Duligall, M.~S. Godfrey, K.~A. Harrison, W.~J. Munroand, and J.~G.  Rarity,
\newblock New J. Phys. {\bf 8}, 249 (2006).

\bibitem{Kok2007}
P.~Kok, W.~J. Munro, K.~Nemoto, T.~C. Ralph, J.~P. Dowling, and G.~J. Milburn,
\newblock Rev. Mod. Phys. {\bf 79}, 135 (2007).

\bibitem{alleaume-2007}
R.~Alleaume, J.~Bouda, C.~Branciard, T.~Debuisschert, M.~Dianati, N.~Gisin, M.~Godfrey, P.~Grangier, T.~Langer, A.~Leverrier, N.~Lutkenhaus, P.~Painchault, M.~Peev, A.~Poppe, T.~Pornin, J.~Rarity, R.~Renner, G.~Ribordy, M.~Riguidel, L.~Salvail, A.~Shields, H.~Weinfurter, and A.~Zeilinger,
\newblock Secoqc white paper on quantum key distribution and cryptography,  2007,
\newblock arXiv.org quant-ph/0701168.

%\bibitem{Vahlbruch2007}
%H.~Vahlbruch, S.~Chelkowski, K.~Danzmann, and R.~Schnabel,
%\newblock New J. Phys. {\bf 9}, 371 (2007).

\bibitem{Boto}
A.~N.~Boto, P.~Kok, D.~S. Abrams, S.~L. Braunstein, C.~P. Williams, and J.~P. Dowling,
\newblock Phys. Rev. Lett. {\bf 85}, 2733 (2000).

\bibitem{Kok}
P.~Kok, S.~L. Braunstein, and J.~P. Dowling,
\newblock J. Opt. B {\bf 61}, S811 (2004).

\bibitem{Migdall}
A.~Migdall, S.~Castelletto, I.~P. Degiovanni, and M.~L. Rastello,
\newblock Appl. Opt. {\bf 41}, 2914 (2002).

\bibitem{Dowling2006}
J.~P. Dowling, A.~Gatti, and A.~Sergienko,
\newblock J. Mod. Opt. {\bf 53}, 573 (2006).

\bibitem{Lloyd}
S.~Lloyd,
\newblock Science {\bf 312}, 1463 (2008).

% \bibitem{Boston}
% M.~B. Nasr, B.~E.~A. Saleh, A.~V. Sergienko, and M.~C. Teich,
% \newblock Phys. Rev. Lett. {\bf 91}, 083601 (2003).

\bibitem{boixo:040403}
S.~Boixo, A.~Datta, M.~J. Davis, S.~T. Flammia, A.~Shaji, and C.~M. Caves
\newblock Phys. Rev. Lett. {\bf 101}, 040403 (2008).

\bibitem{boixo:090401}
S.~Boixo, S.~T. Flammia, C.~M. Caves, and J.~M. Geremia,
\newblock Phys. Rev. Lett. {\bf 98}, 090401 (2007).

\bibitem{PRA1}
B.~P.~Lanyon, M.~Barbieri, M.~P.~Almeida, and A.~G.~White, 
\newblock Phys. Rev. Lett. {\bf 101}, 200501 (2008). 

\bibitem{PRA2}
D.~Gross, S.~T.~Flammia, and J.~Eisert, 
\newblock Phys. Rev. Lett. {\bf 102}, 190501 (2009).

\bibitem{Gottesman99}
D.~Gottesman,
\newblock The heisenberg representation of quantum computers,
\newblock in {\em Group22: Proceedings of the XXII International Colloquium on
  Group Theoretical Methods in Physics}, edited by S.~P.~C. et. al., p.~32,
  Cambridge, MA, International Press, 1999,
\newblock arXiv quant-ph/9807006.

\bibitem{Bartlett02}
S.~D. Bartlett, B.~C. Sanders, S.~L. Braunstein, and K.~Nemoto,
\newblock Phys. Rev. Lett. {\bf 88}, 097904 (2002).

\bibitem{Jozsa2003}
R.~Jozsa and N.~Linden,
\newblock Proc. R. Soc. Lond. A {\bf 459}, 2011 (2003).

\bibitem{Alfano:86}
R.~R. Alfano, Q.~X. Li, T.~Jimbo, J.~T. Manassah, and P.~P. Ho,
\newblock Opt. Lett. {\bf 11}, 626 (1986).

\bibitem{Alexei2}
A.~Gilchrist, G.~J. Milburn, W.~J. Munro, and K.~Nemoto,
\newblock arXiv.org quant-ph/0305167  (2003).

\bibitem{SamCaves94}
S.~L. Braunstein and C.~M. Caves,
\newblock Phys. Rev. Lett. {\bf 72}, 3439 (1994).

\bibitem{general_estimation_theory}
S.~L. Braunstein, C.~M. Caves, and G.~J. Milburn,
\newblock Ann. Phys. {\bf 247}, 135 (1996).

\bibitem{footnote1}
We note that our choice of initial states contain the vacuum and so we must restrict our nonlinear unitary transformation to $k>0$.

\bibitem{footnote2}
Our initial states are not the usual superposition of coherent states that one typically thinks of for use in quantum metrology (they generally have the form $\ket{-\alpha} + \ket{\alpha}$). 
These states have, however, been used as logical qubits in certain quantum computation circuits~\cite{Milburn2002,Jeong2002,Ralph2004,Nature2005}. 
In some aspects they are like the original paradoxical Schrodinger cat state, but they are also different from the state most commonly (but not always) meant by ``cat state'' in the literature.

\bibitem{Ralph2001}
T.~C. Ralph, W.~J. Munro, and G.~J. Milburn,
\newblock arXiv.org quant-ph/0110115  (2001).

\bibitem{master}
S.~Hamaji,\newblock Master's thesis,
\newblock The University of Tokyo, 2006.

%\bibitem{footnote4} We are going to examine the validity of the approximation $e^{i \theta \hat n^k}|\alpha \rangle \sim - |\alpha \rangle$ in more detail. Consider the following overlap $d=\langle \alpha |e^{i \theta \hat n^k}|\alpha \rangle $ with $\theta |\alpha|^{2k} =\pi$. If the approximation is correct then $d=-1$. However, we can calculate $d$ exactly by expanding the coherent states in terms of a number state basis and then applying the operator $e^{i \theta \hat n^k}$ to those number states. Numerically evaluating $d$, with $\alpha=20$ and $k=1/2$ for instance, gives $d=-0.997$ indicating the approximation is pretty good. The approximation of course gets better the larger $\alpha$ is.

\bibitem{Munro}
W.~J. Munro, K.~Nemoto, G.~J. Milburn, and S.~L. Braunstein,
\newblock Phys. Rev. A {\bf 66}, 023819 (2002).

\bibitem{Giovannetti2006}
V.~Giovannetti, S.~Lloyd, and L.~Maccone,
\newblock Phys. Rev. Lett. {\bf 96}, 010401 (2006).

\bibitem{Lee2002}
H.~Lee, P.~Kok, and J.~P. Dowling,
\newblock J. Mod. Opt. {\bf 49}, 2325 (2002).

\bibitem{Milburn2002}
T.~C. Ralph, W.~J. Munro, and G.~J. Milburn,
\newblock Proc. SPIE {\bf 4917}, 1 (2002).

\bibitem{Jeong2002}
H.~Jeong and M.~S. Kim,
\newblock Phys. Rev. A {\bf 65}, 042305 (2002).

\bibitem{Ralph2004}
T.~C. Ralph, A.~Gilchrist, G.~J. Milburn, W.~J. Munro, and S.~Glancy,
\newblock Phys. Rev. A {\bf 68}, 042319 (2003).

\bibitem{Nature2005}
D. Leibfried, E. Knill, S. Seidelin, J. Britton, R.~B. Blakestad, J. Chiaverini, D.~B. Hume, W.~M. Itano, J.~D. Jost, C. Langer, R. Ozeri, R. Reichle, and D.~J. Wineland,
\newblock Nature {\bf 438}, 639 (2005).

\end{thebibliography}
\end{document}